# Unravelling the Power of Single-Pass Look-Ahead in Modern Codecs for Optimized Transcoding Deployment[1]

**Vibhoothi, Julien Zouein, François Pitié, Anil Kokaram**
**Sigmedia Group, Department of Electrical Engineering**
**Trinity College Dublin, Ireland**
vibhootv@tcd.ie, anil.kokaram@tcd.ie

**Abstract** – Modern video encoders have evolved into sophisticated pieces of software in which various coding tools interact with each other. In the past, single-pass encoding was not considered for Video-On-Demand (VOD) use-cases. In this work, we evaluate production-ready encoders for H.264 (x264), H.265 (HEVC), AV1 (SVT-AV1) along with direct comparisons to the latest AV1 encoder inside NVIDIA GPUs (40 series), and AWS Mediaconvert's AV1 implementation. Our experimental results demonstrate single pass encoding inside modern encoder implementations can give us very good quality at reasonable compute cost. The results are presented as three different scenarios targeting High, Medium, Low complexity accounting quality-bitrate-compute load. Finally, a set of recommendations are presented for end-users to help decide which encoder/preset combination might be more suited to their use case.

## 1. Introduction

The ever-increasing demand for online video content has led to the emergence of technologies aimed at reducing transcoding costs in both on-premise and cloud-based environments [1]. In a typical video workflow, which includes transcoding, metadata parsing, and streaming playback with HTTP adaptive streaming (HAS/AS), transcoding consumes a significant share of available resources. Given the increasing volume of video traffic, the resources consumed by video transcoding come under increasing scrutiny. HAS/AS was developed to standardize the way in which different bitrate/quality versions of the same clip could be used to reduce the overall bandwidth requirement for video traffic. But transcoding remains a core part of the creation of those versions therefore the resource issue for transcoding persists. We therefore need encoding algorithms which balance the three core resources: bitrate, quality, and compute.

At the moment it is acceptable industrial lore that the highest quality/lowest bitrate trade-off is only possible through multipass encoding [2] [3] [4]. This incurs substantial computational cost and is not very suitable for live streaming. Because of the increasing complexity of modern codecs (AV1, HEVC, VVC) and the demand for encoders in live broadcast applications [5], more effort has been put into developing optimal single pass encoding scheme [6] [7]. In fact, these schemes involving "lookahead" have evolved significantly, leveraging metadata (e.g., motion information, rate-distortion trade-off, etc.) extracted from frames ahead to inform encoding decisions and the coding process [7] [8]. This results in lower computational cost compared to multipass encoding. It is now widely suspected that these single-pass schemes may be competitive, but no quantitative assessment has been conducted to date. In this paper, we make the following contributions.
1.  **Practical Dataset for Codec Comparison**: We use new high quality source material taken from American Society of Cinematographers (ASC) StEM2 [9] dataset (4k@24fps). Our clips (62) are longer (4 – 30 secs) than the short clips typically used in encoder development (2-4 seconds), and together with this production-ready content provides a more practical assessment of the encoders under test.

[1] This work is supported by ADAPT-SFI Research Center, Ireland with Grant ID 13/RC/2106 P2, and Horizon CL4 2022 - EU Project Emerald – 101119800.

2. **Practical Production-Ready Codecs**: We focus on analysing production-ready encoder implementations instead of a research codebase. We select representative codecs once we have some evidence they are used in production at scale, hence SVT-AV1 is our reference for AV1 since Meta uses SVT-AV1 [10] for Instagram Reels. In addition, we use x264 for the H.264 Standard, x265 for the HEVC standard. For evaluating the hardware encoding performance, we test AV1 encoding in NVIDIA's 40 Series GPU. Lastly, for evaluating cloud performance, we test the latest AV1 implementation inside AWS-MediaConvert. We evaluate single-pass and multi-pass settings at 12 different target bitrates for 5-6 different target presets. We use the VBR control method as typically used in production.
3. **Practical Codec Evaluation**: In evaluations like these, given N clips tested with M codecs each using 5 presets and 12 bitrates, the total number of data points, 62NM, becomes very large (> 35k in our case). To allow us to draw insights from this large amount of data we analyze the data from the perspective of different target requirements. Our evaluation therefore focuses on the following aspects:
   a. The ability of the encoder to retain and achieve the desired target bitrate within certain bitrate boundaries;
   b. Capability to attain perceptually lossless quality; and
   c. The encoding complexity for the dataset. Results are presented for three distinct scenarios, representing high, medium, and low complexity use cases: i) High-Quality Agnostic-Complexity, ii) High-Quality Low-Complexity, and iii) Agnostic-Quality Low-Complexity settings.

   We also briefly analyze the results using Bit-Distortion Aggregation method (Smart BD-Rate) [11] to provide a more practical representation of the dataset within each scenario. The analysis with Smart BD-Rate did not change our findings.

Our findings suggest that in high-complexity 4K encoding scenarios, SVT-AV1 1-pass VBR encoding at preset-2 outperforms all other codecs. Specifically, it achieves approximately 72% bitrate savings compared to x264, 39% compared to x265, 50% compared to nvenc-av1, and 40% compared to AV1 in AWS MediaConvert, as measured by the BD-Rate of video multimedia assessment fusion (VMAF). For the medium-complexity 4K encoding scenario, SVT-AV1 1-pass VBR encoding at preset-6 outperforms all the other codecs. The NVIDIA's AV1 encoder (nvenc-av1) was able to achieve 5% better bitrate savings than "x265-Medium@2-pass" encoding settings. Lastly, when we evaluated the AWS Mediaconvert solution, it was very similar to x265-veryslow 2-pass, at 1.8% BD-Rate loss but 25 times faster.

By examining production-ready systems, we can make recommendations for choosing and parameterising production-ready single-pass and multi-pass workflows. We expect that this work will help to further evolve the development of single-pass encoders.

## 2. Technical Overview of Single-Pass Encoding System

The key challenge for a production-ready encoder is to hit a target bitrate or bitrate range for a piece of encoded video content while ensuring that the output picture quality is sufficiently high. Additionally, it must achieve this at some reasonable computational cost. This intuitive constraint leads to the concept of rate/distortion optimisation (RDO) in which encoder parameters (e.g., quantiser step size, motion search range, block size) are optimised to achieve that balance [12]. In practice we must also distribute encoded bits evenly in some way over the content. This ensures that decoder buffers do not overflow. Intuitively this implies a "bit budget" for encoding a certain number of frames. It is the "rate control" algorithm in a practical deployment of an encoder which is responsible for this clip-level behaviour while again attaining good picture quality. The key intuition behind a good rate controller is to assign bits from the bit budget proportionally to picture complexity in some way. Hence for complex frames (for example) we may deploy more bits than in simpler frames and we assign those bits from our fixed bit budget. But until we visit a frame and analyze the complexity at that time instant, we do not know how to assign a bit budget for that frame.

Therefore, two strategies have broadly emerged. In multi-pass encoding, the encoder encodes a clip multiple times at each iteration learning more about the relative complexity of the frames and updating bit allocations accordingly. In single-pass encoding, the encoder has either to make its best guess (using statistical model-based approaches) as to what might happen in the future or allow a few frames of look-ahead to learn about the near-future. Multi-pass encoding be definition will perform better at RDO than single pass encoding but at significantly higher complexity because of the multiple encoder passes. Modern encoders now have evolved a range of "modes" that encapsulate both RDO and Rate Control, e.g., single-pass, multi-pass (2-pass or 3-pass) and Constant Bitrate (CBR), Variable bitrate (VBR); and Constant Rate factor (CRF).

In the early days, during MPEG-2 development (1999), Mohsenian et al [13] showed that a single-pass encoding using MPEG-2 with buffer constraints can be useful for real-time applications for broadcast and digital applications using CBR and VBR rate-control mode. In 2005, Ma et. al. [14] proposed a technique which is a one-pass rate control at frame level, with a partial two-pass rate control at macroblock level which improved the target bitrate accuracy. Around 2007, Chen et. al. [15] explored various rate-control algorithms and its application. Again, single-pass was recommended for live VOD use-case with VBR, and two-pass rate-control mode used for storage use-cases.

Fast-forward to 2021, Hao et. al. [8] demonstrated that single-pass encoding of AV1 (SVT-AV1) is useful for VOD applications and can achieve similar or better performance than x264/5, VP9, notably the encoder achieves 10-20% better bitrate savings for the same compute complexity. It was noted SVT-AV1 (preset 5) was able to achieve more than 20% bitrate savings compared to VP9 (preset 1). Later in 2021, Nguyen et. al. [16] showed AV1 can outperform HEVC-HM by 11.51% for UHD and FHD test sequences.

The x265 encoder is an open-source implementation of the HEVC standard [17]. This was mainly developed by Multicoreware and supported by VideoLAN. The encoder aims to achieve 40-50% efficiency over its predecessor (x264). The encoder supports multiple rate-control options including 1-pass, 2-pass, multi-pass for CBR, VBR, Constant Quality, and CRF. Within the framework of single pass encoding, the x265 encoder incorporates a lookahead window feature, facilitating the computation information prior to the encoding process. Notable parameters computed during this pre-encoding phase include i) Distortion Costs (e.g., PSNR, SAD, SATD, SSIM) for both Intra and Inter frames, ii) bit information at the coding tree level, iii) average distortion metrics at the block and frame levels, iv) percentages of block types utilized for coding, and v) the positional data of key frames such as last B-frame and I-frame. This pre-encoding data is leveraged during the final encoding phase, thereby reducing the need for redundant computations, and saving CPU cycles. For instance, motion estimates obtained during the lookahead process are directly utilized in the final encoding, thus skipping re-encoding of that specific frame. This reduces computational load substantially. However, based on the encoder preset user choices (veryslow vs veryfast), the computation of a full-set of features for the first-pass or the lookahead can take significant time. In the case of multi-pass encoding, this information along with other video-level statistics are used to make coding decisions in the second or subsequent pass.

The Scalable Video Technology for AV1 (SVT-AV1) is an AV1 codec implementation designed as a production-ready encoder, adopted by the Alliance of Open Media (AOM) Software Implementation Working Group (SIWG). In this work, we considered the latest stable release SVT-AV1 (v.1.8.0, December 2023) [7]. For the single pass encoding mode in the encoder, a lookahead window is defined which computes frame metadata for a certain number of frames ahead (Lookahead Distance: LAD). For a given LAD, two measurements are made, i) Motion Estimation Distortion: computes the sum of absolute difference with the reference frame during motion-estimation, ii) Variance of Motion Estimation Costs for 8x8 blocks across the frame. In the case of multi-pass encoding, the encoder uses this information along with a subset of coding tools in the first pass to improve final video quality in the second pass.

The encoder also contains various parameters to find the correct balance between encoding bitrate, quality, and complexity. These are mediated through preset options. Presently, the encoder has 13 presets (0 to 12), with approximately 35 distinct options fine-tuned across Speed-presets. For instance, different presets determine the maximum number of reference frames allowed. Presets 6 and below all allow up to 7 reference frames and this limit gradually declines to 4 frames in the fastest preset (preset 12). Certain presets/options are available to allow fine-tuning in specific transcoding use-cases. For instance, enabling the Low-Delay configuration is appropriate for screen content coding in video conference applications that could yield a substantial 35% improvement in bitrate savings [18]. For more information on impact of certain coding tools in SVT-AV1, please refer to Appendix A1.

## 3. Experimental Setup

To study the behavior of the encoders, we have selected a cinematic sequence (see Section 3.1, Datasets). For encoding comparison, we chose 3 software and 1 hardware encoder, and 1 cloud-based encoding system (see Section 3.2, Encoder Configuration). For computing the objective metrics, we are using VMAF Library [19] (`8b3b975c`) using CUDA GPU acceleration to compute metrics in real-time (30-40 fps vs 1-2 fps).

### 3.1 Dataset

We use 62 shots (out of 158) from the StEM2 Dataset from American Society of Cinematographers [9]. The original sequence is 17:30 mins long with a 3840x2160p resolution at 24 fps (SDR 10 bits). Each shot has a minimum duration of 4 seconds, with an average duration of 9 seconds. Overall duration of the dataset is 08:37 mins. Figure 1 shows a screenshot of the dataset, and Figure 2 represents the Spatial Energy (SE) and Temporal Energy (TE) of the video computed using the open-source Video Complexity Analyzer (VCA) [20]. SE and TE serve as indicators of video complexity; the SE is computed via a low-pass Discrete Cosine Transform (DCT) across the video, providing a per-frame value, while the TE determined by the maximal temporal variance between consecutive frames. Higher TE values signify increased motion within the video, whereas elevated SE values denote spatially intricate structures, both contributing to increased encoding complexity.

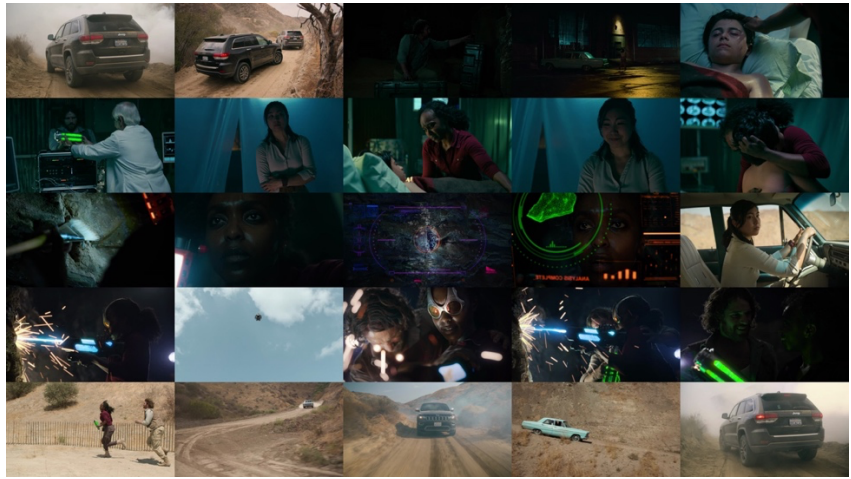

FIGURE 1: SCREENSHOTS OF THE StEM2 DATASET. THE DATASET CONTAINS CINEMATOGRAPHY CONTENT AT 4K@24FPS.

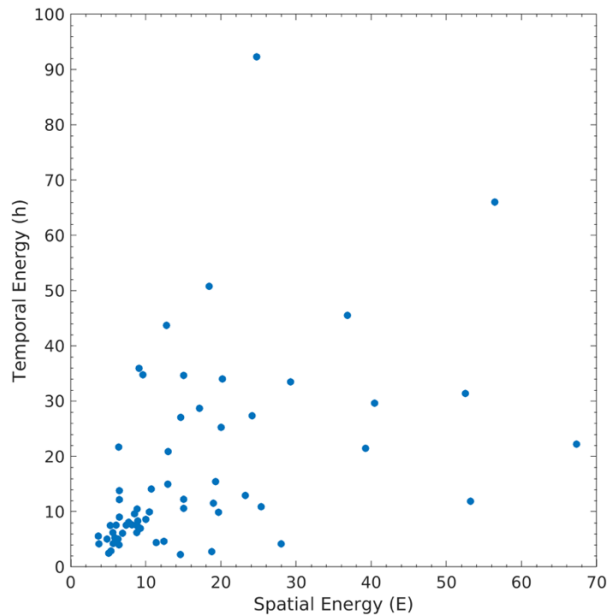

FIGURE 2: DATASET DISTRIBUTION WITH RESPECT TO SPATIAL ENERGY (E, X-AXIS), AND TEMPORAL ENERGY (H, Y-AXIS). CLIPS THAT APPEAR AT THE TOP RIGHT HAVE HIGH CONTENT COMPLEXITY AND THOSE AT THE BOTTOM RIGHT HAVE LOW CONTENT COMPLEXITY. MOST PRODUCTION-READY CLIPS ARE IN THE MID-BAND AREA.

### 3.2 Encoder Configuration

The encoders to be tested are listed in Table 2. Sample command-lines are listed in Appendix A1 for all 5 codecs in the VBR modes. A set of 12 target bitrates is selected for the codecs, {0.5, 1M, 2M, 3M, 4M, 6M, 8M, 10M, 12M, 14M, 16M, 20M}. For each of the target bitrates, we have set the Keyframe interval to be 131, maximum output bitrate to be 120% of the target bitrate, buffersize to be twice of the target bitrate. For example, if the target bitrate is 8M, max rate is set to be 9.6M, buffersize to be 16M. In total we have computed more than 35k encodes (35,268) for analysis. For a particular preset, say x264 veryslow, we have 12 target bitrates * 62 videos resulting in 744 encodes, or for a given video, we are computing and analysing around 5,000 datapoints.

| # | Encoder | Rate Control Options and Passes | Presets Tested | Total Encodes |
|---|---|---|---|---|
| 1 | X264, `Core 164 (LAVC 60.3.100)` | Variable Bitrate (VBR) in 1-pass and 2-pass | veryslow, slow, medium, fast, veryfast, ultrafast (6) | 4464x2; 8928 |
| 2 | X265, *3.5+102-34532bda1* | Variable Bitrate (VBR) in 1-pass and 2-pass | veryslow, slow, medium, fast, veryfast, ultrafast (6) | 4464x2; 8928 |
| 3 | SVT-AV1, `v1.8.0` | Variable Bitrate (VBR) in 1-pass and 2-pass | 2, 4, 6, 8, 10, 12; (6) | 4464x2; 8928 |
| 4 | NVENC-AV1 `Cuda 12.1 (LAVC 60.3.100)` | Variable Bitrate (VBR) in 1-pass and 2-pass | P1, P3, P4, P5, P7; (5) | 3720x2; 7740 |
| 5 | AWS MediaConvert – AV1 | Quality Variable Bitrate Mode (QVBR) | QVBR Level 10 | 744 |

TABLE 2: THE ENCODERS ANALYZED IN THIS WORK. EACH ROW ALSO SHOWS THE HIGH-LEVEL CONFIGURATIONS TESTED. EACH ENCODER/PRESET IS EVALUATED AT 12 DIFFERENT TARGET BITRATES TO CREATE RATE DISTORTION (RD) CURVES.

### 3.2.1 Rationale for Choice of AV1 Preset in AWS MediaConvert

The AWS MediaConvert platform was assessed using the AV1 encoder with Quality Variable Bit Rate (QVBR) as the designated rate-control mode, as AWS currently exclusively supports QVBR for AV1 encoding. The AWS AV1 implementation is less explored in available literature, and direct comparisons to other encoders are limited. Note that 4K and 10-bit support for AV1 was introduced in AWS only in early 2022 [21]. To determine the appropriate QVBR level and quality thresholds applicable, we tested with the "CrowdRun1928x1080" video sequence, because of its demanding high-motion dynamics and intricate scenes. We encoded this clip with10 QVBR levels at 6 different bitrates. Our goal was to determine the range of settings over which it was possible to attain VMAF over 80.

On our testing, it was observed that higher QVBR is required to achieve higher bitrates, and there is an upper-bound of achievable bitrate and quality per QVBR level. Specifically, QVBR level 10 could achieve the full range of desired quality and bitrate, from very low (< 2 Mb/s) to very high (>12 Mb/s), whereas at QVBR level 4, the encoder could not achieve > 70 VMAF score and creates streams with less than 5.8 Mb/s, regardless of the target bitrate.

*Note on AWS MediaConvert Costs*: When the study was conducted in February 2024, the AWS cost for EU (Ireland) region for 4K, AV1 for <=30FPS Single pass HQ encode (Pro Tier) was $1.84099/min. We require to construct 12 bitrates per-shot, and we have 62 shots in our dataset. Average cost for 744 encodes was $1.09 per encode (Min: $0.30, Max: $3.27). Total cost ~$300 including S3 storage.

## 4. Experimental Results: Perceptual Metrics Evaluation

To extract more clear meaning from these experiments, we defined 3 different use-cases for 4K Streaming.
1. *Scenario 1 (S1)*: High-Quality Agnostic-Complexity Encoding: High-Quality Premium 4K Streaming.
2. *Scenario 2 (S2)*: High-Quality Low-Complexity Encoding: High-Quality Normal 4K Streaming.
3. *Scenario 3 (S3)*: Agnostic-Quality Low-Complexity Encoding: Average-Quality 4K Streaming.

For the BD-Rate Measurement, we report *Conventional BD-Rate,* i.e., computing BD-Rate of shots individually and then finding averaging BD-Rate gains for the entire dataset. We also performed an analysis using Smart BD-Rate [8] [11] but this did not change our relative findings.

Figures 4 and 5 summarise the main findings where each of the selected scenarios (S1, S2, S3) is highlighted for a particular preset and codec combination. Figure 4 shows the BD-Rate (%) savings compared to x265-veryslow@2-pass, where lower is better. The (R,D) points are obtained by finding the average across the whole dataset for a particular bitrate and preset combination (see Section 4.4). Figure 5 reports the percentages of videos able to achieve VMAF score greater than 88 at a given preset and codec combination.

### 4.1 Performance of Single-Pass vs Multi-Pass
In all cases, we find in Figure 4 that the one-pass encoding mode (solid line) of every codec performs as well as a two-pass encoding mode (dotted line). In the case of SVT-AV1 for example, the encoding time overhead is around 4-9% when switching to 2-pass, while for x264 and x265, it can be around 20% and 18% respectively. The bitrate savings we obtain with the additional overhead of 2-pass is around 5%. We observe that for low-complexity encoding settings switching to 2-pass can cause encoding complexity to increase up to 50% (e.g., ultrafast in x264, 2.6 vs 5.9 hours). The hardware encoder of NVIIDA is unique, achieving a limited operating range at very low complexity and the gains between 1-pass and 2-pass is around 5%. Lastly, for the AWS AV1, we only considered one preset so we cannot comment exactly on one-pass versus 2-pass, but we note that the AWS solution was able to achieve very similar performance as x265 medium at 2-pass at 26% lower compute requirements.

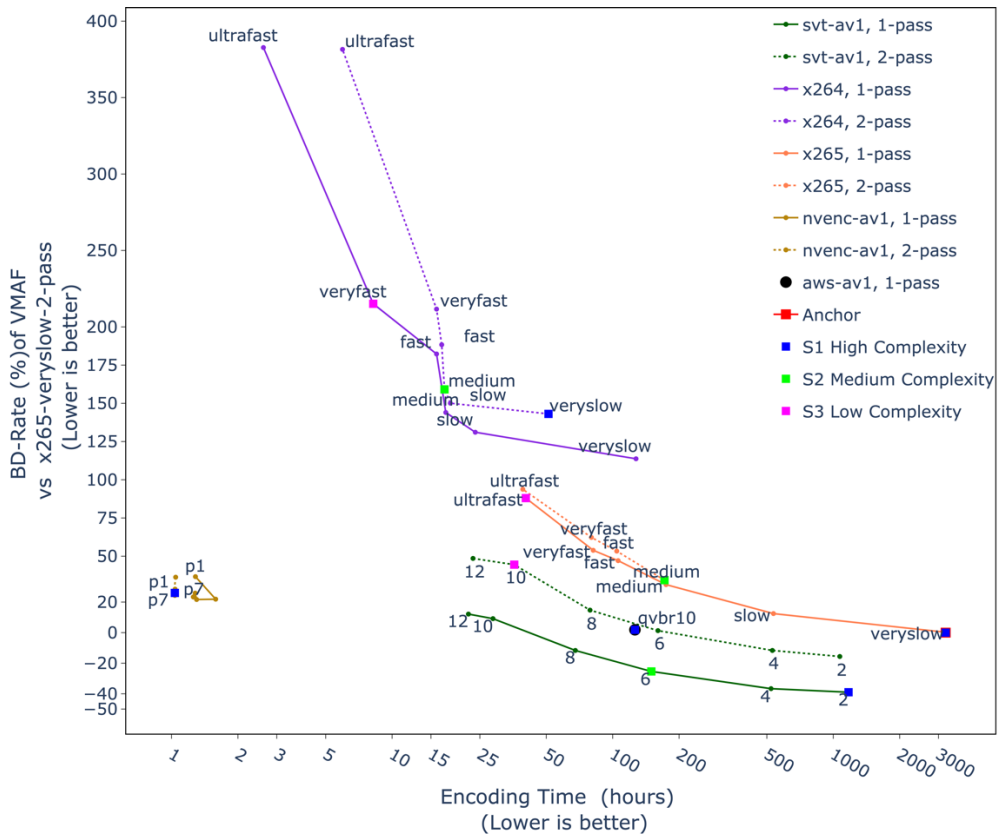

FIGURE 4: BD-RATE (%) DEVIATION (VMAF) VS ENCODING TIME FOR THE ENTIRE DATASET BD-RATE REFERENCE WAS X265-VERYSLOW AT 2-PASS. FOR BOTH AXES, LOWER IS BETTER. FOR THE TESTED CODECS, 1-PASS WAS CONSISTENTLY ABLE TO ACHIEVE HIGHER BITRATE SAVINGS THAN 2-PASS. SVT-AV1@1-PASS WAS ABLE TO ACHIEVE BEST GAINS. BITS AND DISTORTION WAS AVERAGED FOR PARTICULAR BITRATE USING HARMONIC MEAN TO REPRESENT ENTIRE DATASET.

## 4.2 Scenario 1 (S1): High-Quality Agnostic-Complexity Encoding

This scenario aligns with the requirements for high-quality premium 4K streaming. The primary objective is to achieve perceptually excellent quality, measured in terms of VMAF, preferably exceeding or closely approaching 90 for mid-band bitrates (e.g., 4 Mb/s). To identify the codec/preset combinations which achieve good S1 performance, we analyse results with respect to three measurements:
  a. For a given codec's rate-control mode (1-pass or 2-pass) and codec-type, we assess the number of clips achieving a VMAF score greater than 88.
  b. We evaluate the quality achieved at 4 Mb/s and identify the preset that yields the highest quality for the dataset.
  c. We count the videos that exceed 15% of the target bitrates for the respective preset.

We report on the ability to achieve the bitrate constraints (b, c) above in Table A.2. That table shows that excepting NVENC-AV1, all the codecs/presets achieve at least 99% bitrate accuracy. For NVENC and x264 the accuracy is still tolerable at 96%; hence for S1 (b, c) above does not discriminate well between the options.

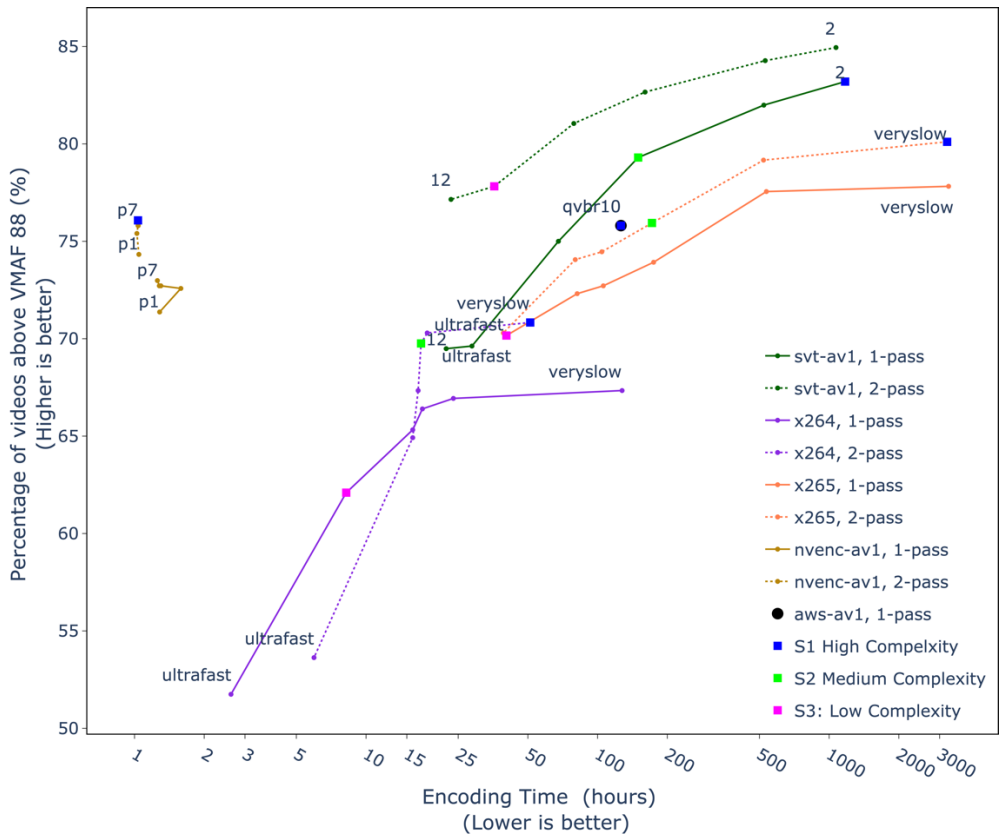

FIGURE 5: ANALYSIS OF DIFFERENT PRESETS IN TERMS OF PERCENT OF THE DATASET ACHIEVING MORE THAN 88 VMAF SCORE (Y-AXIS) COMPARED WITH ENCODING TIME (X-AXIS).

In Figure 5 we can see that, for SVT-AV1, 4 out of 6 presets in 1-pass mode were able to achieve a VMAF of score >88 (desired score) for 75% of dataset. Preset 2 achieved the maximum of 83.2% (619 videos). Hence preset 2 is a good representation for S1 and has a complexity of 1000 hours, which is at the high end as expected. For x265, 2 out of 6 presets in 1-pass achieved the desired score for 75% of the dataset. Switching to 2-pass at veryslow preset achieved the maximum of 80.1% for x265. For x264, using 2-pass can achieve bigger fraction of dataset satisfying criterion (a). Hence veryslow at 2-pass for x264 and x265 is a good representation for S1. The x265 is particularly highly complex with 3233 hours of compute hours. For the AWS-AV1, the QVBR10 preset of AWS-AV1 was able to get 75.8% (564) of videos above the desired score. Lastly for NVENC, in 2-pass mode we achieve maximum number of clips satisfying (a) with 76.07%. The encoding complexity between 1-pass and 2-pass for NVENC is very similar (1.25 hours vs 1.04 hours). Although S1 is complexity-agnostic, NVENC was with lowest complexity.

The preset that best represents S1 over our codecs, are shown with a "blue" dot in Figures 4 and 5. Overall, we see that SVT-AV1 achieves the best compromise between BD-Rate savings and criterion (a) above using 1-pass at preset 2.

### 4.3 Scenario 2 (S2): High-Quality Low-Complexity Encoding

This scenario is tailored for high-quality 4K streaming while also considering encoding time as a crucial factor in determining the optimal (Rate, Distortion) points. The preferred settings in this scenario prioritize achieving the highest VMAF for the fastest encoding time at the given bitrate. To identify the codec/preset combinations which achieve good S2 performance, we use two measurements.
   a. Use the S1 presets as baseline to compare the BD-Rate (%) performance.
   b. Compare the encoding time reduction with respect to S1.

In Figure 4 we can see that for SVT-AV1, preset 6 was able to achieve a BD-Rate (%) gain of 25% over best x265 preset (veryslow at 2-pass). This has a loss of 21% compared to S1 preset of SVT-AV1. This resulted in more than 85% reduction in encoding complexity (1070 hours vs 149.7 hours). Hence, we recommend Preset-6 at 1-pass. For x264, migrating to medium preset (at 2-pass) results in a 5.16% BD-Rate (%) loss while reducing encoding complexity by 66.2% (51.6 vs 17.2 hours). For x265, switching to medium preset (2-pass) results in a 94% decrease in encoding complexity (3233 vs 171 hours), with a ~27% BD-Rate loss over S1 preset. Hence, for x264 and x265, we recommend medium preset for S2 scenario. In the case of NVENC hardware solution, encoding time differences between different presets are negligible, with all gains falling within measurement errors (0.5-2%). The measured BD-Rate gains between 1-pass mode and 2-pass mode are around 3-5%, while for transition from the slowest preset (Preset 7, higher preset means slower) to faster preset (p2/p1 for e.g.), we obtain BD-Rate gains between 0.6 to 13%. As the preset behavior was not strictly monotonic, coupled (Golden Yellow line in Figure 4) with insignificant time difference between presets, we recommend keeping with slowest encoder preset for reliable results. Lastly, as AWS-AV1 solution lacks alternative rate-control modes in AWS, we recommend adhering to the currently available mode.

The presets which represent best for Scenario 2 is highlighted with "Green dot" in Figure 4 and 5. Again, we observe that SVT-AV1 was able to achieve best BD-Rate savings along with addressing criterion (a) in our analysis for S2. The bitrate analysis for the selected preset is shown in Table A.3.

## 4.4 Scenario 3 (S3): Agnostic-Quality Low-Complexity Encoding

This scenario addresses average quality 4K streaming requirements, where minimizing bitrate is crucial, and target quality can be comparatively lower, while considering encoding time as a key factor. This aligns with typical User-Generated Content (UGC) grade 4K compression or non-premium/mobile 4K streaming. To identify the codec/preset combinations which achieve good S3 performance, we only use one measurement, that is, target the encoding time to be under 40 hours for the entire dataset.

We can see that from the Figure 4 and 5, in SVT-AV1 transitioning to Preset 10 with 2-pass encoding significantly reduces encoding complexity to be around 35.8 hours. The rationale for choosing a 2-pass mode for SVT-AV1 here is the encoding overhead is around 20% for an 8% increased VMAF>88 coverage for our dataset (69.6% to 77.8%). Thus, Preset 10 at 2-pass is the choice for S3. For x265, in the fastest speed-preset in 1-pass (ultrafast), we achieved desired VMAF score coverage for 70% of dataset, this was attained at 40.3 hours. For x264, choosing the veryfast preset at 1-pass achieved 62% of dataset coverage with just 8 hours. When we compare the performances to S1, we would observe large BD-Rate (%) loss like 73% for SVT-AV1. However, if we analyze the performance of SVT-AV1 and X265 S3 presets, they surpass X265-medium@2-pass and X264-veryslow@2-pass by approximately 3% respectively. Thus, for Scenario 3, we choose them.

As it was observed in the past [22], visual perception vastly differs based on a display device's size (mobile phone vs television). This scenario would be particularly useful for mobile streaming use-cases. And techniques like multi-codec streaming [10] [23], could be viable to use in this method, where modern codecs like AV1 could be used for low-bitrates scenarios. Figure 6 below demonstrates the BD-Rate comparisons between these selected codecs for the 3 different scenarios.

**<u>Note on Smart-BD-Rate</u>**: When have also evaluated the codec performances by separately summing the {Rate, Distortion} pairs [8], [11] and then determining the BD-Rate, over computing BD-Rate individually for the dataset and averaging. Following [8] we used Harmonic Mean for calculating the results. We observe similar trends in performance, and the recommended presets are similar. See Appendix A3 for more information.

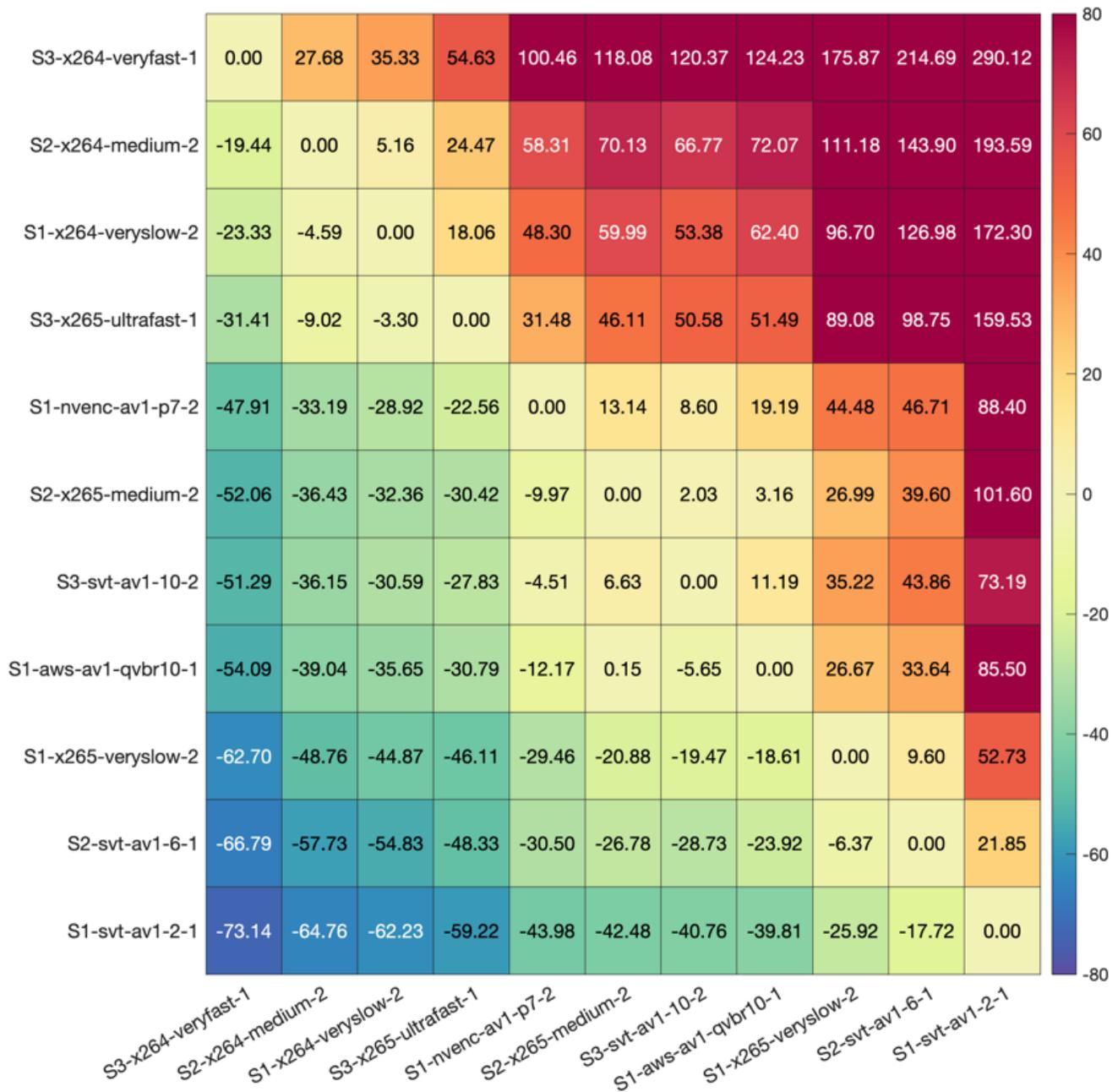

FIGURE 6: THIS FIGURE ILLUSTRATES THE BD-RATE (%) COMPARISON GRID AMONG VARIOUS CODEC PRESETS UTILIZED IN THE THREE DIFFERENT SCENARIOS, REPRESENTING HIGH (S1), MEDIUM (S2), AND LOW (S3) COMPLEXITIES. NEGATIVE VALUES INDICATE IMPROVEMENT. THE X-AXIS REPRESENTS THE ANCHOR, WHILE THE Y-AXIS REPRESENTS THE TEST CONFIGURATIONS. FOR EXAMPLE, TRANSITIONING FROM X264-VERYSLOW-2 PRESET AT 1 PASS TO SVT-AV1'S PRESET 2 AT 1-PASS YIELDS A 62.23% BITRATE SAVINGS.

The RDcurve results for selected codecs across 3 scenarios is depicted in Figure 8 below. The bitrate constraints for S3 are shown in Table A.4.

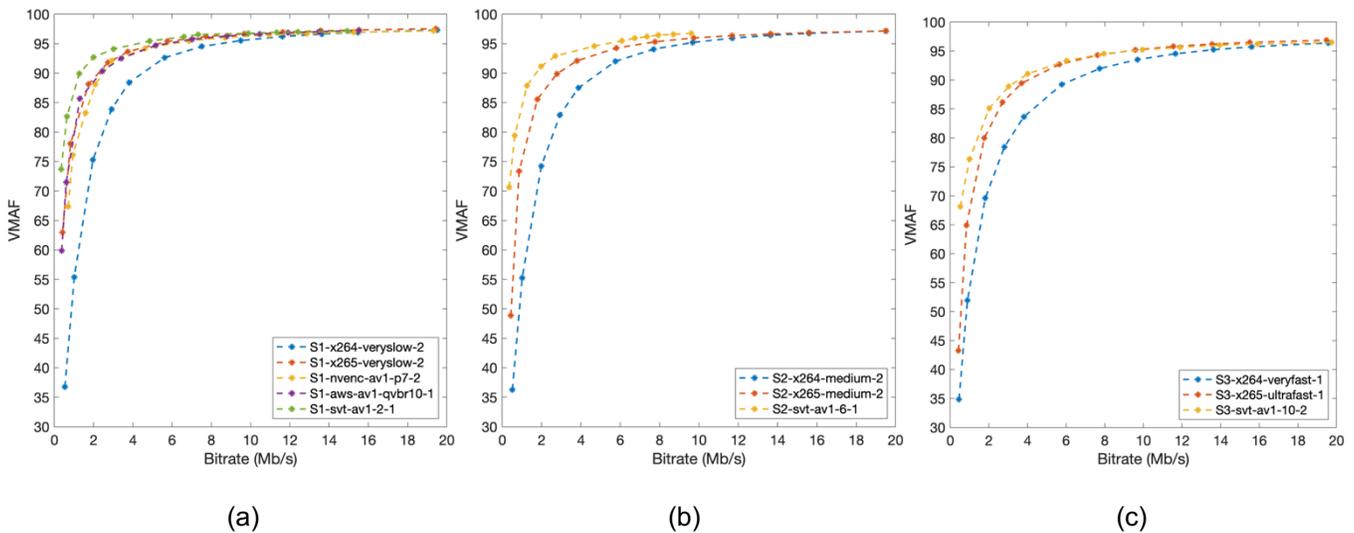

(a)  (b)  (c)

FIGURE 8: RATE-DISTORTION (RD) CURVES FOR THREE DIFFERENT SCENARIOS (S1/S2/S3) FOR THE StEM2 DATASET. S1 CORRESPONDS TO HIGH-COMPLEXITY, S2 DENOTES MEDIUM COMPLEXITY, S3 DENOTES LOW COMPLEXITY CODING SETTINGS. THE VALUES ARE OBTAINED BY COMPUTING HARMONIC MEAN FROM 62 INDIVIDUAL (R,D) POINTS OF DIFFERENT SHOTS. X-AXIS IS BITRATE IN MB/S, Y-AXIS IS VMAF. WE WOULD PREFER CODECS WITH LOWEST-BITRATE AND HIGHEST-QUALITY.

## 5. Experimental Results: Computational Load Evaluation

To evaluate the compute load of the encoding systems, we consider two different approaches. Figure 8 shows the sum of the encode time required for all different shots for a particular preset, reported for the 11 different codecs for 3 different presets. From this distribution, we can see that x265 "veryslow" preset (2-pass) has the highest complexity, followed by SVT-AV1-Preset 2 (1-pass). We can also observe that each scenario which we defined reduces the encode complexity by half or more.
Figure 9 shows the second analysis, computing encoding percentage gains/loss. This is computed by keeping a particular preset as an anchor and computing the encode-time difference on a per-preset basis and computing the mean for the single clip. We can see that certain combination can result in massive increases in encoding complexity, while others massively decrease encoding complexity. With this result, in conjunction with Smart-BDR, we can choose the desired switch of codec/preset to migrate to improved bitrate/quality/compute cycle trade-off.

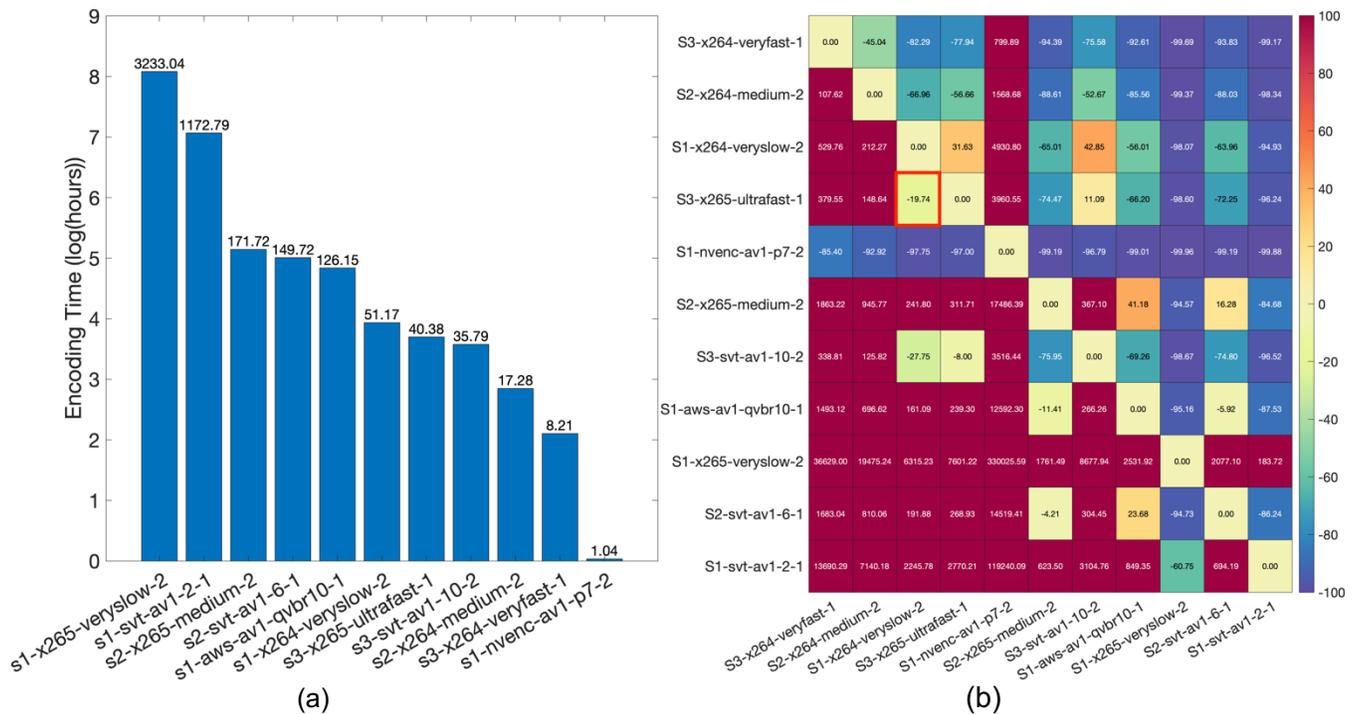

FIGURE 8: LEFT (a): ENCODING TIME FOR DIFFERENT CODECS OVER THE THREE SCENARIOS (S1, S2, S3), X-AXIS DENOTES DIFFERENT CODECS BEING USED, AND Y-AXIS DENOTE THE TOTAL ENCODING TIME MEASURED IN LOG(HOURS) REQUIRED FOR THE ENTIRE DATASET FOR THAT SCENARIO (12 BITRATES * 62 CLIPS). THE ABSOLUTE ENCODING TIME IS SHOWN AT THE TOP OF EACH BAR. RIGHT (b): ENCODE TIME COMPARISON HEAT MAP FOR DIFFERENT CODECS/SCENARIOS. X-AXIS AND Y-AXIS DENOTE DIFFERENT ENCODERS BEING TESTED. FOR INSTANCE, IF WE MIGRATE FROM S1-x264-VERY-SLOW-2 TO S3-x265-ULTRAFAST-1 WE WOULD HAVE A 20% FASTER PIPELINE (HIGHLIGHTED WITH RED-BLOCK).

## 6. Recommendations

While the above-described scenarios are useful, in practice a user is more likely to require an answer to the question "if I switch encoders, what gains/losses would I incur?" The notes below summarise a few takeaways that help to answer this question.

- SVT-AV1 outperforms every other codec in terms of bitrate-quality trade-off.
- We observe that single-pass can be competitive to 2-pass in most scenarios. And we achieve an additional ~5% bitrate savings when switching to 2-pass.
- If you are using x264 at any preset or any 2-pass settings (focusing on VBR here), switching to any other choice including NVIDIA AV1 Hardware encoder or AWS Mediaconvert, or x265/SVT-AV1, can yield a substantial BD-Rate savings.
- AWS Mediaconvert's AV1 implementation gives us similar performance as SVT-AV1's preset 6 in 2-pass, or x265-veryslow at 2-pass. Migrating to AWS AV1 solution can reduce encoding complexity by around 25x. However, the encoding cost is not cheap, so the switch to AWS based solution should be made based on business requirements.
- NVIDIA's AV1 encoder gives around 5% better BD-Rate gains over x265@Medium using 2-pass. The encoding complexity reduction is about 165x (171 hours vs 1 hour). At any given point the NVIDIA AV1 encoding is faster than any x264 settings by more than 46%.
- For a finer-quality switch, deploying a multi-codec streaming [23] approach can be beneficial to cover a wider range of bitrate savings. That is, for high-complexity scenarios choosing SVT-AV1 at Preset 2 (S1) can give 38% bitrate savings with 33% reduction in the encoding complexity of x265 veryslow @2-pass (3233 vs 1172 hours). For preset 6 in SVT-AV1, we can get 55% BD-Rate (%, VMAF) improvement over x265-medium at 2-pass with 10% lower encode-complexity.

# 7. Conclusion

The work presents a comparison of various codecs on a practical dataset like ASC StEM 2. We have shown results based on 3 standards (H.264, H.265, and AV1). We considered x264, x265, SVT-AV1, NVIDIA's latest AV1 hardware implementation, and AWS Mediaconvert's AV1 offering. Experimental results demonstrate that we can find a single-pass encoding mode which can perform similar to a 2-pass encoding mode. We analyzed the results with three different scenarios, targeting i) high-quality agnostic-complexity scenario (S1), ii) agnostic-quality and low-complexity use-case (S2), iii) agnostic-quality and agnostic-complexity use-case (S3). The results showed that one-pass encoding with SVT-AV1 across different presets in 1-pass can achieve 35% bitrate savings over x265 at 2-pass for similar or ~30% lower complexity. Overall, for different codecs, we measured the impact of 2-pass to be around 5-8% over 1-pass, while encode time for fast presets can be increased up to 50%. The NVIDIA AV1 Encoder can be competitive with x264 (at least 46%) and x265 (5% for medium@2-pass) with a marginal encoding cost. The AWS AV1 encoding solution at 1-pass (QVBR10) can give similar performance as x265-veryslow in 2-pass. These experimental results suggest that switching encoders in production and transcoding environment can give noticeable improvements in scale.

## Appendix A1: Impact of Encoding Parameters in Single-pass Encoding in SVT-AV1

To examine the impact of available features for single-pass settings, we selected 9 different coding tools/features as listed in Appendix A2 which we can tune. We used "SteM2_18266-18579.y4m" video (4K@24fps, with 313 frames), where a systematic tool-off test is analyzed (disabling one feature at a time). Each of these settings was encoded at nine different bitrates (500 kb/s to 10 Mb/s) in 1-pass VBR Mode at Default preset settings (Preset 10). Table A.1 presents the results of the test. The bitrate and VMAF metric values are reported for target bitrate of 4000 kb/s. The bitrate savings is reported using Bjontegaard Delta (BD-Rate) [24] where the default encoder is the anchor. We are using the AOM-CTC [25] implementation of BD-Rate. We can see that disabling some tools can give us a BD-Rate gains up to 12%, while for other cases this can include a loss of 1%. However, such outcomes are inherently content-dependent, necessitating diversified clip evaluations aligned with specific transcoding use-cases.

| # | Tool Off | Bitrate (Kb/s) | VMAF | PSNR-Y (dB) | BD-Rate (%) VMAF |
|---|---|---|---|---|---|
| 1 | Default | 3479.402 | 91.495 | 50.606 | 0.000 |
| 2 | --enable-dlf 0 | 3506.605 | 91.428 | 50.590 | 1.192 |
| 3 | --enable-cdef 0 | 3488.512 | 91.862 | 50.478 | -7.227 |
| 4 | --enable-restoration 1 | 3442.836 | 91.462 | 50.595 | -0.264 |
| 5 | --enable-tpl-la 0 | 3424.716 | 91.479 | 50.600 | -0.171 |
| 6 | --enable-mfmv 1 | 3438.263 | 91.548 | 50.596 | -0.830 |
| 7 | --enable-dg 0 | 3402.444 | 91.454 | 50.581 | -0.216 |
| 8 | --fast-decode 1 | 3439.587 | 91.594 | 50.622 | -2.832 |
| 9 | --enable-tf 0 | 3574.181 | 92.187 | 50.728 | -12.729 |
| 10 | --enable-overlays 1 | 3400.891 | 91.470 | 50.582 | -0.167 |

TABLE A.1: THE EFFECT OF DISABLING VARIOUS CODING TOOLS IN A TOOL-OFF SIMULATION. THE VIDEO, "stem2_sdr_rec709_420f_2160p24_178_10bit_18266-18579.y4m" (4K@24FPS), IS ENCODED USING SVT-AV1 WITH 9 DIFFERENT TARGET BITRATES@1-PASS VBR MODE. THE VMAF AND MEASURED BITRATE AT 4M IS REPORTED ALONG WITH BD-RATE (%) GAINS USING VMAF (NEGATIVE IS BETTER).

## Appendix A2: Command-line Options for Encoding

*X264*
1-pass VBR
```
ffmpeg -y -I $input.y4m -g 131 -keyint_min 131-b:v $tbr -maxrate $tbr*1.5 -
bufsize $tbr*2 -c:v libx264 -threads 1 -preset $preset -tune psnr -x264-
params scenecut=0 -f mp4 $output.mp4
```

2-pass VBR
```
ffmpeg -y -i $input.y4m -g 131 -keyint_min 131 -b:v $tbr -maxrate $tbr*1.5 -
bufsize $tbr*2 -c:v libx264 -threads 1 -preset $preset -tune psnr -pass 1 -
passlogfile $passfile.log -x264-params scenecut=0 -f mp4 /dev/null && ffmpeg
-y -i $input  -g  131 -keyint_min 131 -b:v $tbr -maxrate $tbr*1.5 -bufsize
$tbr*2 -c:v libx264 -threads 1 -preset $preset -tune psnr -pass 2 -
passlogfile $passfile.log -x264-params scenecut=0 $output.mp4
```

*X265*
1-pass VBR
```
ffmpeg -y -I $input.y4m -g 131 -keyint_min 131-b:v $tbr -maxrate $tbr*1.5 -
bufsize $tbr*2 -c:v libx265 -threads 1 -preset $preset -tune psnr -x265-
params scenecut=0 -f mp4 $output.mp4
```

2-pass VBR
```
ffmpeg -y -i $input.y4m -g 131 -keyint_min 131 -b:v $tbr -maxrate $tbr*1.5 -
bufsize $tbr*2 -c:v libx265 -threads 1 -preset $preset -tune psnr -pass 1 -
passlogfile $passfile.log -x265-params scenecut=0 -f mp4 /dev/null && ffmpeg
-y -i $input  -g  131 -keyint_min 131 -b:v $tbr -maxrate $tbr*1.5 -bufsize
$tbr*2 -c:v libx265 -threads 1 -preset $preset -tune psnr -pass 2 -
passlogfile $passfile.log -x265-params scenecut=0 $output.mp4
```

*SVT-AV1*
1-pass VBR
```
./SvtAv1EncApp -i $input.y4m --keyint 131 --tbr $tbr -lp 1 --rc 1 --preset $preset -b $output.mp4
```

2-pass VBR:
```
./SvtAv1EncApp -i $input.y4m --keyint 131 --tbr $tbr -lp 1 --rc 1 --passes 2 --preset $preset -b $output.mp4
```

*NVENC-AV1*
1-pass VBR
```
ffmpeg -y -i $input.y4m -g 131 -keyint_min 131 -b:v $tbr -maxrate $tbr*1.5 -bufsize $tbr*2 -c:v av1_nvenc -rc vbr -threads 1 -preset $preset -no-scenecut 1 $out.mp4
```

2-pass VBR
```
ffmpeg -y -i $input.y4m -g 131 -keyint_min 131 -b:v $tbr -maxrate $tbr*1.5 -bufsize $tbr*2 -c:v av1_nvenc -rc vbr -threads 1 -preset $preset -no-scenecut 1 -multipass 2 $out.mp4
```

## Appendix A3: Tabular Results for Scenarios

### Scenario 1 (S1): High-Quality Agnostic-Complexity Encoding: High-Quality Premium 4K Streaming.

| # | Codec | No of Videos with Bitrate Overshoot > 15% from TBR | Videos with VMAF > 88 | | Encoding Time for the whole dataset (hours) |
| | | | All TBR (744) | 4000kb/s (62) | |
|---|---|---|---|---|---|
| 1 | SVT-AV1@1-pass – Preset 2 | **3** | **619** | **58** | **1172.78** |
| 2 | X264@2-pass – VerySlow | 8 | 527 | 43 | 51.18 |
| 3 | X265@2-pass-Veryslow | 2 | 596 | 54 | 3233.04 |
| 4 | NVENC-AV1@2-pass-Preset 7* | 28 | 566 | 54 | 1.25 |
| 5 | AWS-AV1@1-pass-QVBR10 | 1 | 564 | 53 | 126.1483 |

TABLE A2: THE HIGH-LEVEL PRESET SELECTION FOR THE SCENARIO 1. NOTE THAT THE NVENC-AV1 PRESET 7* MEANS THE HIGH COMPLEXITY SLOW ENCODE.

### Scenario 2 (S2): High-Quality Low-Complexity Encoding: High-Quality Normal 4K Streaming.

| # | Codec | No of Videos with Bitrate Overshoot > 15% from TBR | Videos with VMAF > 88 | | Encoding Time reduction (%) from S1 Preset choice | Encoding Time for the whole dataset (hours) |
| | | | All clips (744) | @4000kb/s (62 clips) | | |
|---|---|---|---|---|---|---|

| # | Codec | No of Videos with Bitrate Overshoot > 15% from TBR | Videos with VMAF > 88 | | VMAF | Encoding Time for the whole dataset (hours) |
|---|---|---|---|---|---|---|
| 1 | SVT-AV1@1-pass – Preset 6 | **3** | **590** | **53** | 86.24 | 149.72 |
| 2 | X264@2-pass – Medium | 2 | 519 | 40 | 66.2 | 17.28 |
| 3 | X265@2-pass-Medium | 3 | 565 | 45 | 94 | 171.71 |

TABLE A3: THE HIGH-LEVEL PRESET SELECTION FOR THE SCENARIO 2.

**Scenario 3 (S3): Agnostic-Quality Low-Complexity Encoding: Average-Quality 4K Streaming.**

| # | Codec | No of Videos with Bitrate Overshoot > 15% from TBR | Videos with VMAF > 88 | | Encoding Time reduction (%) from S1 Preset choice | Encoding Time for the whole dataset (hours) |
|---|---|---|---|---|---|---|
| | | | All TBR (744) | 4000kb/s (62) | | |
| 1 | SVT-AV1@2-pass – Preset10 | 8 | 579 | 51 | 96.94 | 35.79 |
| 2 | X264@1-pass – Veryfast | 0 | 462 | 37 | 83.95 | 8.21 |
| 3 | X265@1-pass-Ultrafast | 3 | 522 | 44 | 98% | 40.38 |

TABLE A4: THE PRESET SELECTION USED FOR DIFFERENT CODECS FOR THE SCENARIO 3.

## Appendix A4: Smart-BD-Rate, a BD-Rate Computation Using Bits-Distortion Averaging Method for Multi-Shot Datasets.

As it was noted in Wu et. al. [8] [11], for videos comprising multiple shots, a more precise calculation of bitrate savings can be achieved by separately summing the {Rate, Distortion} pairs and then determining the BD-Rate. This means that, we would need to average the values to represent for one (R,D) point in the curve, we use harmonic mean. With this, we can recompute BD-Rate with the new averaged values for the given scenario (presets and codec combination).

Figure A1 shows the (R,D) curve averaging for a preset. Figure A2 reports BD-Rate (%) gains comparison for all codecs. Compared to regular BD-Rate (%), for some cases there is noticeable shift in observed gains. For example, in the Classic-BDR computation, the gains from migrating from AWS-AV1 to SVT-AV1 at Preset-6@2-Pass were to be 23.83%, while here it is only having a loss of 0.04%. Similarly, on migrating from x264's veryslow@2-Pass to SVT-AV1 at Preset-2@1-Pass, the BD-Rate (%) gains have increased from 62.23% to 72.38%. However, the trend of gains and codec-switch remains the same.

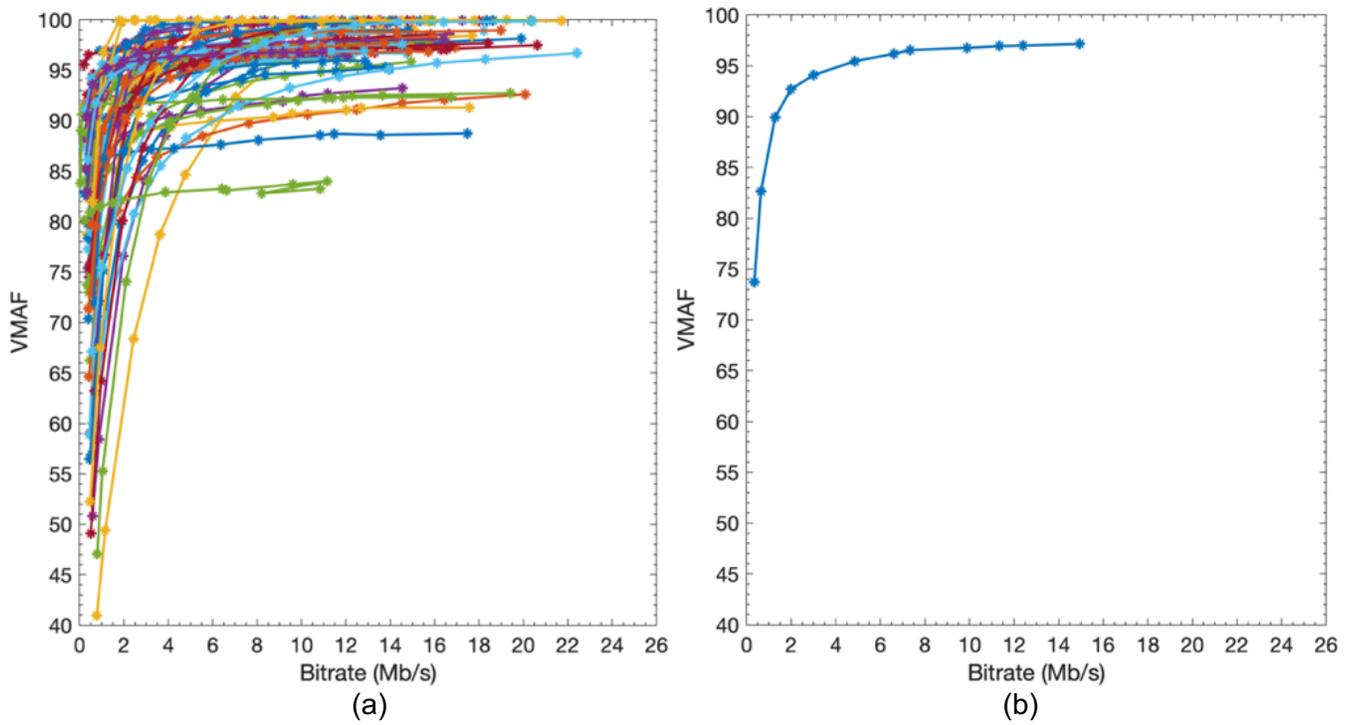

FIGURE A1: THE (R,D) CURVE FOR SVT-AV1 AT PRESETS 2 USING 1-PASS FOR THE DATASET. THE X-AXIS DENOTES BITRATE IN MB/S, AND Y-AXIS DENOTE VMAF SCORE. ON THE LEFT (a), (R, D) CURVES FOR 62 VIDEOS IS SHOWN. RIGHT (b): THE NEW (R,D) CURVE OBTAINED BY HARMONIC MEAN AT THIS SCENARIO. IT'S IMPORTANT TO NOTE THAT THE HARMONIC MEAN MAY INTRODUCE BIAS TOWARDS HIGH-COMPLEXITY (HIGH-BITRATE) VIDEOS.

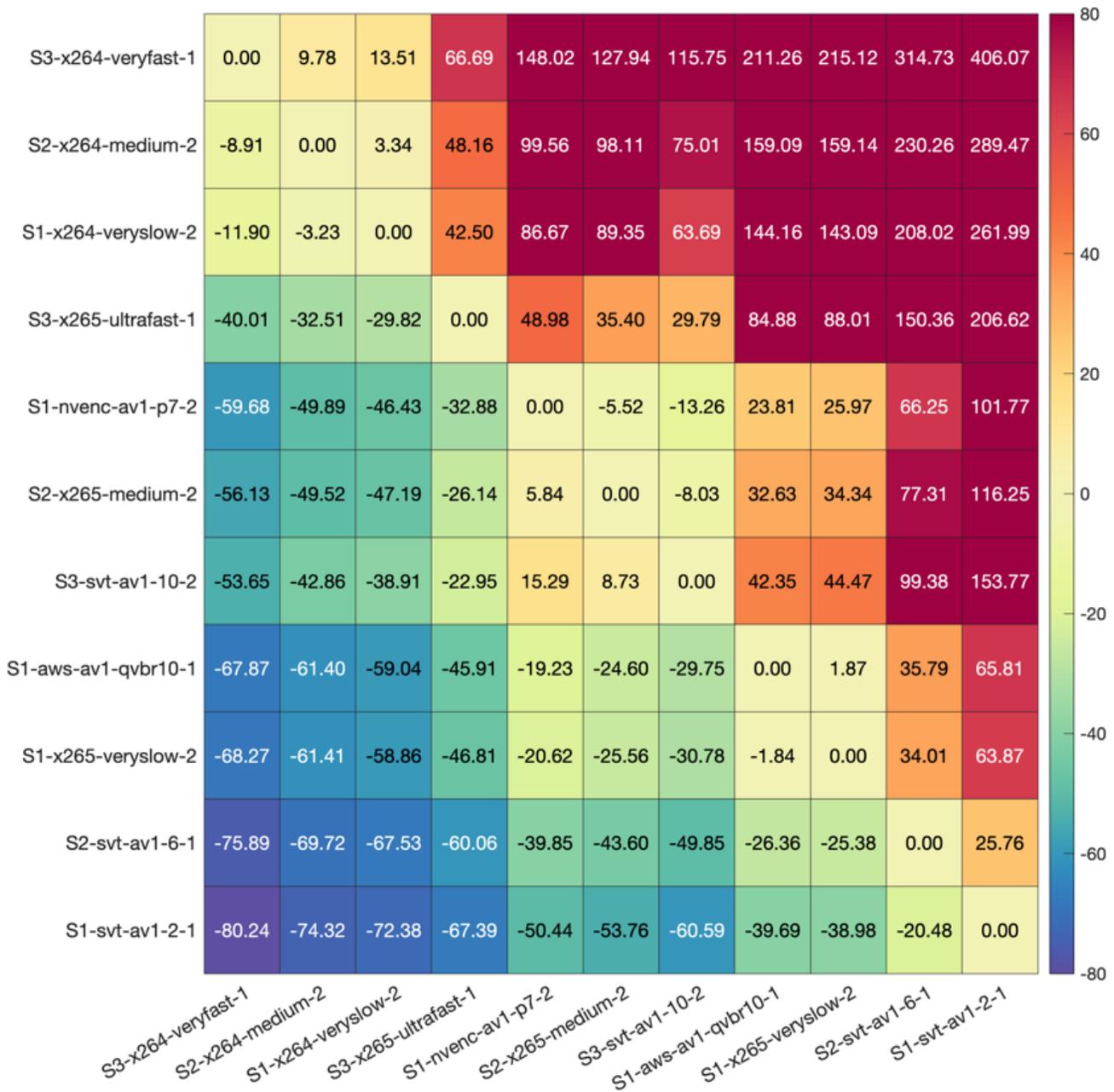

FIGURE A2: THIS FIGURE ILLUSTRATES THE BD-RATE (%) COMPARISON GRID AMONG VARIOUS CODEC PRESETS UTILIZED IN THE THREE DIFFERENT SCENARIOS, REPRESENTING HIGH (S1), MEDIUM (S2), AND LOW (S3) COMPLEXITIES. THE (R,D) POINTS ARE OBTAINED BY COMPUTING HARMONIC MEAN FOR A PARTICULAR BITRATE FOR A CODEC FOR THE ENTIRE DATASET OF INDIVIDUAL SHOTS. NEGATIVE VALUES INDICATE IMPROVEMENT. FOR EXAMPLE, TRANSITIONING FROM X264-VERY-LOW-2 PRESET AT 1 PASS TO SVT-AV1'S PRESET 2 AT 1-PASS YIELDS A 72.38% BITRATE SAVINGS. THE X-AXIS REPRESENTS THE ANCHOR, WHILE THE Y-AXIS REPRESENTS THE TEST CONFIGURATIONS. NOTE THAT HARMONIC MEAN CAN CAUSE A BIAS TOWARDS HIGHER BITRATE VALUES.